\documentclass[amsmath,amssymb,superscriptaddress,nobalancelastpage,prb,twocolumn]{revtex4}

\usepackage{graphicx}
\usepackage{varioref}
\usepackage{xr-hyper}
\usepackage{xcolor}
\definecolor{bl}{rgb}{0.0,0.2,0.6}
\usepackage{hyperref}

\def\Re{\textrm{Re}}
\def\Im{\textrm{Im}}

\def\ImSt{\Sigma^{\prime\prime}}
\def\ReSt{\Sigma^{\prime}}

\newcommand{\LSCOov}{La$_{1.77}$Sr$_{0.23}$CuO$_4$ }

\begin{document}

 \title{Nodal Landau Fermi-Liquid Quasiparticles in Overdoped La$_{1.77}$Sr$_{0.23}$CuO$_4$}

  \author{ C. G. Fatuzzo       }
 \affiliation{Institute for Condensed Matter Physics, \'{E}cole Polytechnique Fed\'{e}rale
 de Lausanne (EPFL), CH-1015 Lausanne, Switzerland} 

\author{Y.~Sassa}
\affiliation{Laboratory for Solid State Physics, ETH Z\"{u}rich, CH-8093 Z\"{u}rich, Switzerland}

\author{M.~M\aa nsson}
\affiliation{Laboratory for Solid State Physics, ETH Z\"{u}rich, CH-8093 Z\"{u}rich, Switzerland}
\affiliation{KTH Royal Institute of Technology, Materials Physics, S-164 40 Kista, Sweden }

\author{S.~Pailh\`{e}s}
\affiliation{Laboratory for Neutron Scattering, Paul Scherrer Institut, CH-5232 Villigen, Switzerland}
\affiliation{LPMCN, CNRS, UMR 5586,
Universit\'{e} Lyon 1, F-69622 Villeurbanne, France}

\author{O.~J.~Lipscombe}
\affiliation{H.\ H.\ Wills Physics Laboratory, University of Bristol, Bristol, BS8 1TL, United Kingdom}
\author{S.~M.~Hayden}
\affiliation{H.\ H.\ Wills Physics Laboratory, University of Bristol, Bristol, BS8 1TL, United Kingdom}

\author{L.~Patthey }
\affiliation{Swiss Light Source, Paul Scherrer Institut, CH-5232 Villigen PSI, Switzerland}

\author{M.~Shi}
\affiliation{Swiss Light Source, Paul Scherrer Institut, CH-5232 Villigen PSI, Switzerland}

\author{M. Grioni}
\affiliation{Institute for Condensed Matter Physics, \'{E}cole Polytechnique Fed\'{e}rale
de Lausanne (EPFL), CH-1015 Lausanne, Switzerland}

\author{H. M. R\o{}nnow}
\affiliation{Institute for Condensed Matter Physics, \'{E}cole Polytechnique Fed\'{e}rale
de Lausanne (EPFL), CH-1015 Lausanne, Switzerland}

\author{J.~Mesot}
\affiliation{Institute for Condensed Matter Physics, \'{E}cole Polytechnique Fed\'{e}rale
de Lausanne (EPFL), CH-1015 Lausanne, Switzerland}
\affiliation{Laboratory for Solid State Physics, ETH Z\"{u}rich, CH-8093 Z\"{u}rich, Switzerland}
\affiliation{Laboratory for Neutron Scattering, Paul Scherrer Institut, CH-5232 Villigen, Switzerland}

\author{O.~Tjernberg}
\affiliation{KTH Royal Institute of Technology, Materials Physics, S-164 40 Kista, Sweden }

  \author{ J.~Chang         }
 \affiliation{Institute for Condensed Matter Physics, \'{E}cole Polytechnique Fed\'{e}rale
 de Lausanne (EPFL), CH-1015 Lausanne, Switzerland}
 \affiliation{Swiss Light Source, Paul Scherrer Institut, CH-5232 Villigen PSI, Switzerland}
 \affiliation{Laboratory for Neutron Scattering, Paul Scherrer Institut, CH-5232 Villigen, Switzerland}
  
\begin{abstract}
Nodal angle resolved photoemission spectra taken on overdoped La$_{1.77}$Sr$_{0.23}$CuO$_4$ are presented and analyzed.
It is proven that the low-energy excitations are true Landau Fermi-liquid quasiparticles. 
  We show that momentum and energy distribution 
 curves can be analyzed self-consistently without quantitative knowledge of the bare band dispersion. 
Finally, 
by imposing Kramers-Kronig 
consistency on the  self-energy $\Sigma$, insight into the quasiparticle residue is gained.  We conclude by comparing our results to 
quasiparticle properties extracted from thermodynamic, magneto-resistance, and high-field quantum oscillation experiments on 
overdoped Tl$_2$Ba$_2$CuO$_{6+\delta}$. 
\end{abstract}

\maketitle
\section{Introduction}
The extent to which Landau Fermi-liquid theory, and its concept 
of quasiparticles \cite{Pines}, applies to the normal state of cuprates is
still under debate~\cite{MarelPNAC13,NevenPNAC13}. Evidence for Landau Fermi-liquid quasiparticles
has been reported by resistivity experiments on highly overdoped La$_{2-x}$Sr$_x$CuO$_4$ (LSCO)~\cite{NakamaePRB03}. 
More recently, unambiguous proof 
has been given by high-field quantum oscillation experiments~\cite{DoironLeyraud07a,VignolleNat08,Sebastian12a,Vignolle11a,NevenNP13}
on both overdoped and underdoped cuprates.
It is, however, puzzling that  no evidence of Landau Fermi-liquid quasiparticle excitations
 has been found from angle resolved photoemission spectroscopy (ARPES),
that is a direct probe of the Green's function: $-(1/\pi)\textrm{Im}G(k,\omega)$~\cite{DamascelliRMP03}. 
Although the ``quasiparticles''  terminology  is widely used to describe the 
excitations of the photoemission spectra, it has never been proven by ARPES 
that the low-energy excitations in the cuprates are indeed true Landau Fermi-liquid
quasiparticles.  
A direct spectroscopic proof of true
Landau Fermi-liquid quasiparticles in the cuprates 
is therefore important.

This paper has two main objectives. 
The first is to prove that the nodal excitations observed in overdoped LSCO by ARPES
are genuine Landau Fermi-liquid quasiparticles. The second is to 
discuss the nodal bare band velocity, $v_b$, and the nodal quasiparticle residue $Z\equiv (1-\partial \ReSt / \partial \omega)^{-1}$,
where $\Sigma$ is the self-energy. 
Perhaps the most compelling spectroscopic proof of Landau quasiparticles 
is the demonstration of a low-energy self-energy that has (1) the form $\ImSt \propto i \omega^2$~\cite{JackoNATPHYS2009}
 and (2) $-Z\ImSt <|\omega|$~\cite{GeorgesPRL2011,GeorgesPRL2013}. 
Proving this requires full knowledge about $\Sigma$ and insight into the bare band $\varepsilon_b$, that is not straightforward to derive 
from an APRES spectrum~\cite{KordyukPRB05}.
Here, we however present an experimental case where
$Z\ImSt$ can be evaluated without quantitative knowledge of $\varepsilon_b$.
In this specific case, it is therefore possible to prove the existence of true Landau Fermi-liquid 
quasiparticle excitations.

\begin{figure*}
\begin{center}
\includegraphics[width=0.75\textwidth]{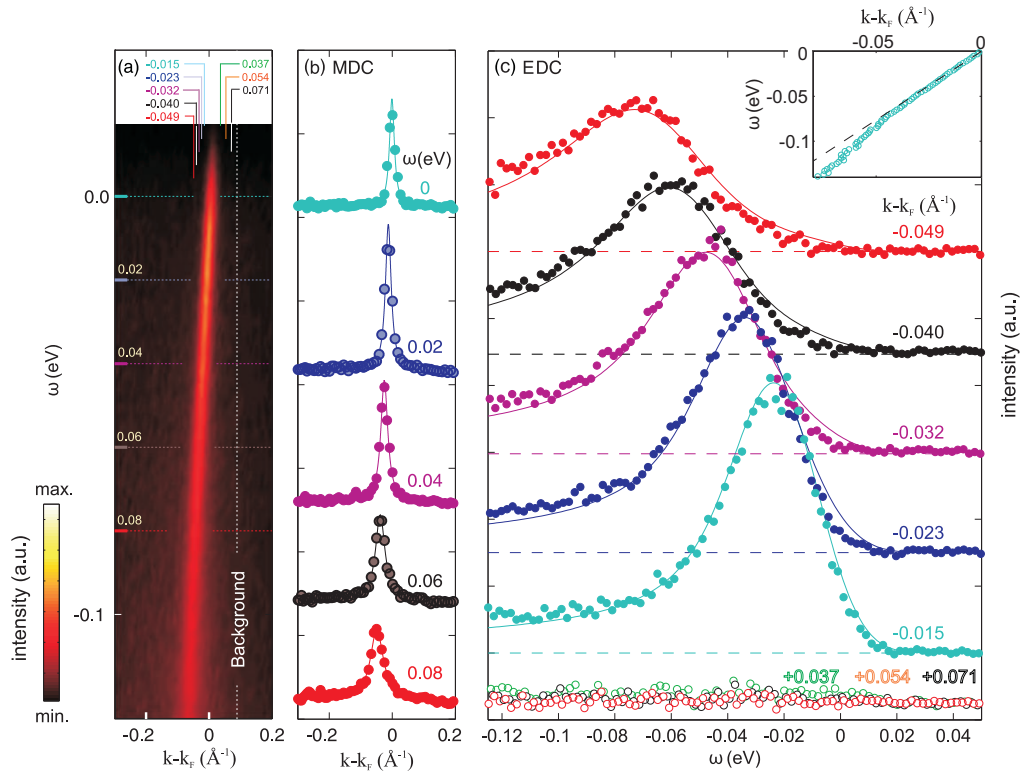}
\end{center}
\caption{(a) Nodal ARPES spectra recorded from overdoped LSCO ($x=0.23$), at $T=15$ K with 55 eV photon. The intensity, displayed 
versus momentum $k-k_F$ (horizontal) and excitation energy $\omega$ (vertical), has a false color scale with white as the most intense as indicated by the 
colorbar. 
(b) Momentum distribution curves (MDCs) of the spectra shown in (a), for fixed energies as indicated. Solid lines
are Lorentzian fits to the data. (c) Energy distribution curves (EDCs) recorded at momenta as indicated. 
A $\omega$-dependent background defined by the EDC at $k-k_F=0.089$~\AA$^{-1}$  (indicated by the vertical white dashed line in (a)) 
has been subtracted. Solid lines display 
the $\omega$-dependence of Eq.~3, multiplied with the Fermi-Dirac distribution and convoluted with the 
instrumental resolution~\cite{noteRES}.
For the sake of visibility, data in (b) and (c) are arbitrarily shifted in the vertical direction. The insert 
of (c) displays the excitation dispersion derived from MDC analysis of the spectra in (a). }
\label{fig:fig2}
\end{figure*}

\begin{figure}
 \begin{center}
\includegraphics[width=0.4\textwidth]{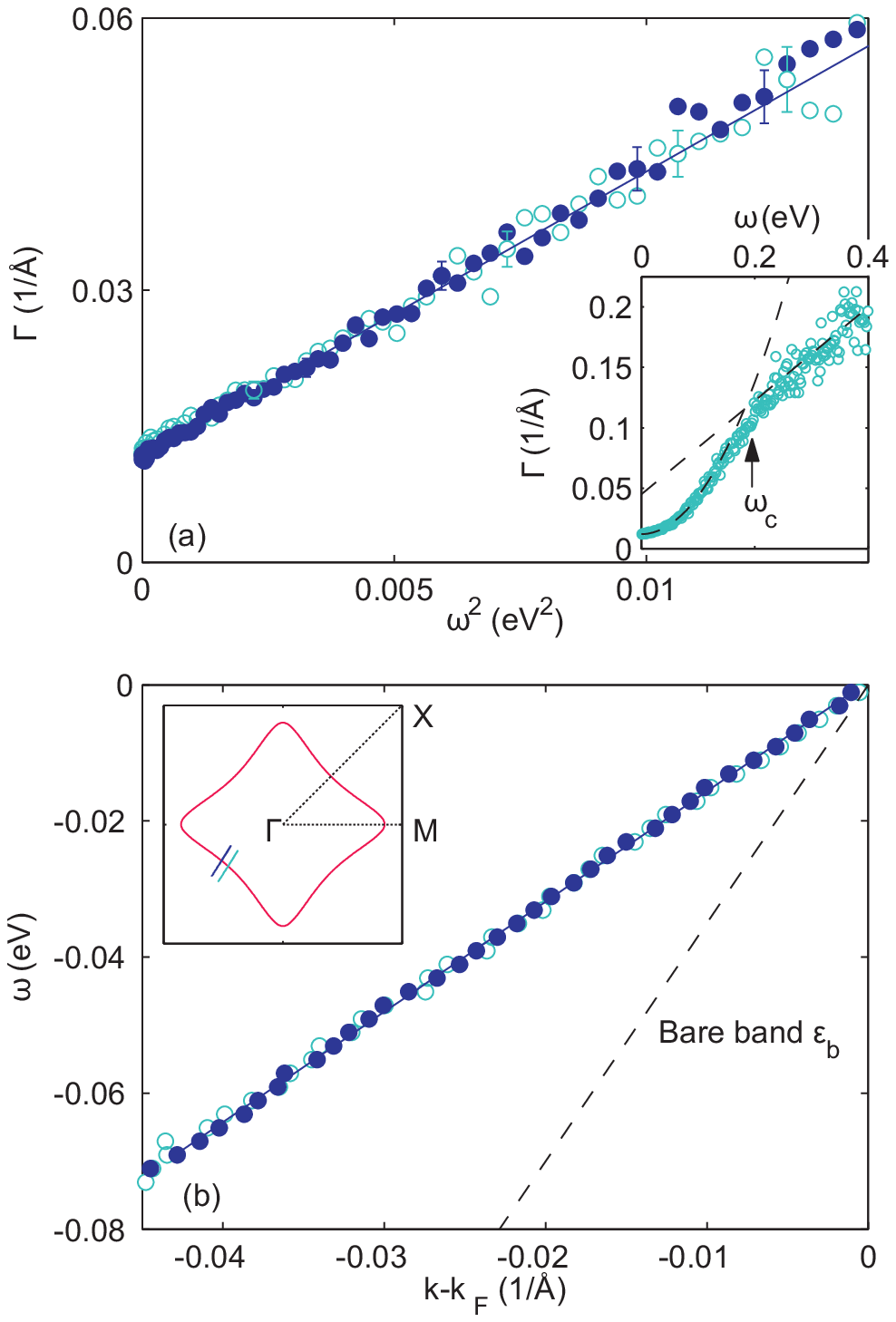}
  \end{center}
  \caption{(a) Linewidth $\Gamma$ of the momentum distribution curves plotted versus excitation energies squared ($\omega^2$), for the two cuts 
shown in the inset of (b). Inset of (a) shows $\Gamma$ vs $\omega$. The deviation from $\omega^2$ dependence defines the energy scale 
$\omega_c$ \cite{ChangNatComm13} -- indicated by the arrow.
(b) Nodal dispersions extracted from MDC analysis for the two cuts shown in the inset. The dashed line indicates the bare-band 
extracted by assuming Kramers-Kronig consistency of the self-energy $\Sigma$, see text. The inset shows the Fermi surface of LSCO $x=0.23$
and the two cuts along which the spectra were recorded.  
 }
  \label{fig:fig2}
  \end{figure}

\section{Methods}
Nodal ARPES spectra of 
overdoped  \LSCOov\ ($T_c=25$K)~\cite{LipscombePRL07,ChangPRB2012} were recorded at the surface 
and interface spectroscopy (SIS) beam line~\cite{SIS} of the Swiss Light Source (SLS) at 
the Paul Scherrer Institute, Switzerland. 
High quality nodal spectra were obtained after 
cleaving~\cite{cleaver} at $T=15$~K under ultra-high vacuum conditions ($p\sim$10$^{-11}$ mbar).
Using 55 eV circular polarized
photons and a SCIENTA 2002 electron analyzer, angular and energy
resolutions corresponding to 0.15$^{\circ}$ (FWHM) and $\sigma=9$~meV (standard Gaussian deviation) 
were achieved. A detailed description of the 
 experimental conditions can be found in Ref.~\onlinecite{ChangNatComm13}.

\section{Results}
Fig.~1(a) shows a colormap -- $I$ vs $(k,\omega)$ -- of ARPES spectra recorded close to the nodal direction 
of overdoped La$_{1.77}$Sr$_{0.23}$CuO$_4$. A selection of corresponding momentum  distribution curves (MDC)
and energy distribution curves (EDC) are displayed in Fig.~1(b,c). 
We start by discussing the MDCs. As this paper focuses entirely 
on the low-energy excitations, MDCs are only 
shown up to the energy scale (80~meV) of the nodal kink shown in the inset of Fig. 1(c).  
In this energy interval, 
the MDC line shapes are symmetric peaks on a constant background.
Therefore, data at constant $\omega$ were analyzed using a Lorentzian function 
$I_0\Gamma/[(\omega-\varepsilon_k)^2+\Gamma^2]$, 
where $\Gamma$ is the linewidth (Fig.~2a), $\varepsilon_k$ is the peak position (Fig.~2b), and $I_0$ is an amplitude (Fig.~3).
The observed nodal excitations disperse with a Fermi velocity $v_F=1.62(2)$~eV\AA~[Fig.~2(b)], consistently 
with previous reports on LSCO~\cite{ZhouNAT03,YoshidaJCMP07}.
The half-width half-maximum, $\Gamma$, is plotted as a function of excitation energy squared 
$\omega^2$ in Fig.~2(a). We find that, for $\omega<\omega_c=0.18\pm0.2$~eV, the linewidth is well 
described by $\Gamma=\Gamma(0)+\eta\omega^2$ 
with $\eta=3.14(4)$~eV$^{-2}$\AA$^{-1}$, and $\Gamma(0)=0.0117(1)$~\AA$^{-1}$. 
The elastic scattering $\Gamma(0)$ is lower
than what is usually reported for LSCO~\cite{YoshidaJCMP07,changPRB08}.
As impurity scattering is one source of elastic scattering~\cite{VarmaPNAS00}, 
low values of $\Gamma(0)$ may be an indication of 
high sample quality.

To reveal the intrinsic physical line shape, a background 
has been subtracted from the EDCs shown in Fig.~1(c). The energy dependent 
background was extracted from the spectra using an energy distribution curve 
on the un-occupied side of the dispersion (indicated by a vertical dashed line in Fig.~1(a)).
This is a common procedure~\cite{GweonPRL11} and an example of a raw background spectrum can be 
found in the supplement of Ref.~\onlinecite{ChangNatComm13}. 
In this fashion, EDCs recorded at a momentum $|k|$ larger than the Fermi momentum $|k_F|$ [displayed with open circles in Fig.~1(c)]
are featureless, demonstrating 
the successful background subtraction. On the other hand, EDCs with $k<k_F$ [full circles] 
reveal the intrinsic line shape of the excitations.

\section{Discussion}
The measured ARPES
intensities $I(k,\omega)$ can be modelled by a product 
of the spectral function $A(k,\omega)=-(1/\pi)\textrm{Im}G(k,\omega)$, a matrix element $M(k,\omega)$, and the
Fermi distribution $f(\omega)$~\cite{DamascelliRMP03}.
Matrix elements typically vary weakly
as a function of $(k,\omega)$. 
 Notice that the excitations 
shown in Fig.~1(b) disperse over less than 10 percent of the Brillouin zone.
It is therefore not unreasonable to ignore matrix element effects. 
In that case,
the ARPES intensity becomes a direct measure
of the occupied part of the spectral function. 
It is common practice to separate the spectral function 
into coherent and incoherent parts, {\it i.e.} $A(k,\omega)=A_{coh}(k,\omega)+A_{inc}(k,\omega)$~\cite{DamascelliRMP03}.
 The coherent part can be written as:
\begin{equation}
  A_{coh}(k,\omega)=\frac{-1}{\pi}\frac{\ImSt(k,\omega)}{(\omega-\ReSt(k,\omega)-\varepsilon_b)^2+\ImSt(k,\omega)^2}
\end{equation}
where $\varepsilon_b$ is the \textit{a priori} unknown bare band, 
and the self-energy  must obey $|\ReSt|\gg|\ImSt|$~\cite{varmaPR02}. 
Experimentally, one would associate sharp dispersing features 
to the coherent part of the spectral function and featureless 
weight to the incoherent part. 
Here we focus on the low-energy part of the spectra, where the 
coherent spectral weight is dominating.

To make progress, 
two justified assumptions are made. 
First, it is assumed that the experimentally unknown bare band 
can be linearized ($\varepsilon_b\simeq v_b (k-k_F)$) near the Fermi level.
Already the observed normalized band (extracted from the MDC analysis) can, to a very good 
approximation, be described by $\varepsilon_k=v_F(k-k_F)$, with $v_F=1.62$~eV\AA~[Fig.~2(b)].
The bare band is expected to have an even larger band velocity --
LDA calculations suggest for example $v_b\simeq3.5$ eV\AA~\cite{AndersenPRL01}. For the excitation 
energies discussed here, curvature effects of the bare band are therefore 
expected to be negligibly small. 
Secondly, we assume that the self-energy, $\Sigma$, is locally momentum independent. 
Globally this assumption is not correct -- the self-energy varies strongly 
as one approaches the anti-nodal region~\cite{ChangNatComm13}. However, locally,  in the vicinity of the 
nodal region, this is a good approximation. As shown in Fig. 2, both the 
band velocity and MDC linewidth are essentially identical for the two different 
nodal cuts. Another indication that $\Sigma$ is momentum independent stems
from the symmetric MDC lineshape shown in Fig. 1b. 
A $k$-dependence of $\Sigma$ along the cut-direction would 
lead to an asymmetric lineshape. As this is not observed, it is concluded that $\Sigma$
is locally independent of momentum both along and perpendicular to the cut direction.

It is thus possible to rewrite the coherent part of the spectral function as:
\begin{equation}
 A_{coh}(k,\omega)=\frac{-1}{\pi}\frac{\ImSt(\omega)}{(\omega-\ReSt(\omega)-v_b(k-k_F))^2+\ImSt(\omega)^2}.
\end{equation}
Notice that this is nothing else than a Lorentzian function in momentum space,
with half-width half-maximum $\Gamma$ given by $\Gamma(\omega)=-\ImSt(\omega)/v_b$.
Experimentally, it is found that $\Gamma\propto\omega^2$ (see Fig. 2). 
Therefore, consistently with true Fermi liquid quasiparticle excitations, 
we conclude that $\ImSt\propto\omega^2$.

A Kramers-Kronig consistent self-energy with $\ImSt\propto \omega^2$
has $\ReSt\simeq -\gamma \omega$ in the low-energy limit.
The unknown constant $\gamma$ is sometimes referred to as the quasiparticle renormalization factor~\cite{GeckPRL07}.
If consistency between MDC and EDC
 poles (dispersions) is enforced,  then
$1/(1+\gamma)=v_F/v_b = Z$. 
The coherent spectral function  can consequently be re-written as: 
\begin{equation}
 A_{coh}(k,\omega)=\frac{Z}{\pi}\frac{v_F\Gamma}{(\omega-v_F(k-k_F))^2+(v_F\Gamma)^2},
\end{equation}
where both $v_F$ and $\Gamma$ are known from the MDC analysis. 
The only unknown parameter, $v_b$, is a prefactor. 
It is therefore possible to model the EDC lineshape without quantitative knowledge of the 
bare band $\epsilon_b$, and with the peak amplitude as the only free parameter -- see solid lines in Fig.~1(c). 
In the displayed energy interval, a consistent description of both EDCs and MDCs 
were obtained from  $A_{coh}(k,\omega)$.

Because $Z=v_F/v_b$ and $\ImSt =-\eta v_b\omega^2$, the product $Z\ImSt = -v_F \eta \omega^2$
can be evaluated without quantitative knowledge of the bare band velocity.
The condition for coherent quasiparticle excitations is 
$-Z\ImSt<|\omega|$~\cite{GeorgesPRL2011,GeorgesPRL2013}.
Using the experimental values of $v_F$ and $\eta$, we find that 
Landau quasiparticles are coherent for $\omega< 1 / v_F \eta \sim 0.19$~eV.
This energy scale is comparable to $\omega_c$ -- the energy scale 
below which $\ImSt\propto \omega^2$ -- and hence re-enforces the interpretation of $\omega_c$ as
an energy scale related to the break down of Landau Fermi-liquid quasiparticle excitations~\cite{ChangNatComm13}.

Finally, we discuss the Kramers-Kronig
relation between $\ReSt$ and $\ImSt$: 
\begin{align}
  \ReSt &= \frac{\mathcal{P}}{\pi} \int_{-\omega_c}^{\omega_c} \frac{\ImSt(\omega^{\prime})}{\omega^\prime-\omega} d\omega^\prime \pm \frac{\mathcal{P}}{\pi} 
\int_{\pm\omega_c}^{\pm W} \frac{\ImSt(\omega^{\prime})}{\omega^\prime-\omega} d\omega^\prime\\\nonumber
&=\ReSt_{qp}+\ReSt_{nqp}
\end{align}
where $\mathcal{P}$ is the principal value and $W$ is the band width. 
To first order, the quasiparticle part yields $\ReSt_{qp}\simeq\gamma_{qp}\omega$ 
where $\gamma_{qp}=2v_b\eta\omega_c / \pi$.
To gain insight into $\ReSt_{nqp}$, we define
$Z_i(\omega)=(1-\partial \Re\Sigma_i / \partial \omega)^{-1}$ so that $Z(\omega)=Z_{qp}(\omega)+Z_{nqp}(\omega)$.
As $Z(\omega)\sim I_0(\omega)$ varies weakly with excitation energies 
(see Fig. 3), 
we infer that 
$Z_{nqp}$ (to first order) is $\omega$-independent .
Hence $\Re\Sigma_{nqp}(\omega)=\gamma_{nqp}\omega$ and $Z=1/(1+\gamma_{qp}+\gamma_{nqp})$.  
As long as the detailed high-energy part of $\Im\Sigma(\omega)$ is 
unknown, it is not possible to directly extract $\ReSt_{nqp} = \gamma_{nqp}\omega$.
This is known as the "tail" problem~\cite{KordyukPRB05}. The linear $\omega$-dependence 
at high-energies, shown in the inset of Fig. 2, yields 
$\gamma_{nqp}\sim  \ln(C/\omega_c)$ where $C$ is an unknown constant.
Hence $\gamma_{nqp}$ diverges only 
logaritmically in the limit $\omega_c\rightarrow 0$~\cite{ChangNatComm13}.
On the other hand, for large $\omega_c$ the role of $\gamma_{nqp}$
will be less important. As $\omega_c=0.18$~eV is a large energy scale, corresponding to a
temperature scale of the order 1000 K, we hypothesize that 
$\gamma_{nqp} \ll 1$. In that case, 
$Z\simeq1/(1+\gamma_{qp})=v_F/v_b$ and hence $v_b =\pi v_F/(\pi-2\eta\omega_c v_F)= 3.8$~\AA.
 This is consistent with the nodal 
 LDA Fermi velocity $v_{LDA} = 3.5$~eV\AA~\cite{AndersenPRL01} calculated for LSCO and 
with values of $v_b$ derived from a numeric self-consistent 
 method~\cite{KordyukPRB05}. The consistent values of $v_b$
further support the conjecture that $\gamma_{nqp} \ll 1$.

\begin{figure}
 \begin{center}
\includegraphics[width=0.4\textwidth]{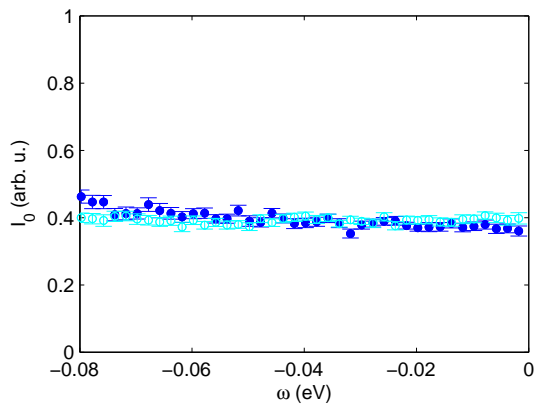}
  \end{center}
  \caption{ Amplitude $I_0$, in arbitrary units, versus excitation energy $\omega$
fot the two cuts in the inset of Fig.~2b. 
Correction by the Fermi-Dirac distribution only influence the data points near 
the Fermi level.}
  \label{fig:fig2}
  \end{figure}

The quasiparticle mass is given by $m_b/m^*=Z \hat{Z}$, where $m_b$ is the bare mass
and $\hat{Z}= 1 + (m_b/\hbar^2k_F) \partial \Sigma^{\prime}(k,0)/\partial k$~\cite{RourkeNJP10,varmaPR02}.
Since the self-energy is locally independent of momentum, the nodal 
quasiparticle mass is given by $m^* = m_b /Z \simeq 2.4 m_b $. 
This is comparable to the momentum averaged values  $m^*\simeq3m_b$
extracted from quantum oscillation~\cite{VignolleNat08,RourkeNJP10} and electronic specific heat experiments 
on overdoped Tl$_2$Ba$_2$CuO$_{6+\delta}$ (Tl2201)~\cite{LoramPHYSC94}.
Remarkably, a Fermi-liquid cut-off energy scale $\omega_c \sim 0.2$~eV was extracted~\cite{McKenziePRL11,McKenziePRB12} 
from angle-dependent magneto-resistance measurements on overdoped Tl2201~\cite{HusseyNatPhys06}. 
This is in good agreement with nodal ARPES spectra recorded on 
LSCO $x=0.23$~\cite{ChangNatComm13}. 
On LSCO, no quantum oscillation or angle-dependent magneto-resistance experiments exist.
Insight into the average quasiparticle mass of overdoped LSCO stems, therefore, alone
from specific heat measurements~\cite{MomonoPHYSICA94}. Compared to Tl2201~\cite{LoramPHYSC94}, a somewhat larger Sommerfeld constant $\gamma_{el}\simeq 12$~mJ/(mole K$^2$) 
is found for overdoped LSCO $x\simeq 0.23$~\cite{MomonoPHYSICA94}, suggesting a larger average quasiparticle mass. 
This is not necessarily inconsistent with the ARPES data. The Fermi-liquid cut-off 
energy scale, $\omega_c$, softens rapidly as a function of Fermi surface angle, and 
the quasiparticle scattering is globally dependent on momentum~\cite{ChangNatComm13}. 
This implies (1) 
that the contribution from non-Fermi liquid excitations will become increasingly 
important and (2) that $\hat{Z}<1$ on certain portions of the Fermi surface. 
Both effects would lead to larger quasiparticle masses.

\section{Conclusions}
In summary, we have proven that the nodal single particle excitations
observed by ARPES in overdoped LSCO are indeed true Landau 
Fermi liquid quasiparticle excitations. 
This result, together with consistent MDC and EDC analysis, 
was obtained without knowing the exact bare band. 
From  Kramers-Kronig consistency of the 
quasiparticle self-energy $\Sigma$,  insight into 
the bare band $\epsilon_b$ and 
the real part of the self-energy $\ReSt$ were obtained. 
An estimate of the nodal quasiparticle residue 
$Z=0.42(7)$ allowed comparison to quasiparticle masses
obtained from thermodynamic and high-field quantum oscillation 
experiments on overdoped Tl$_2$Ba$_2$CuO$_{6+\delta}$ compounds~\cite{VignolleNat08}.

{\it Acknowledgments:} This work was supported by the Swiss NSF 
(through NCCR, MaNEP, and grant Nos. 200020-105151 and PZ00P2-142434), the Ministry
 of Education and Science of Japan, and the Swedish Research Council.
  The experimental work was performed at the SLS of the Paul Scherrer Institut,
  Villigen PSI, Switzerland. We thank the X09LA beamline~\cite{SIS} staff
   for their technical support.


\begin{thebibliography}{31}
\bibitem{noteRES}
 There are at least two ways to mimic the effect of instrumental resolution. 
 One is to use the measured $\Gamma_0$ and the width of the Fermi step that already include instrumental resolution.
 Another method is to use $\Gamma_0(intrinsic)=(\Gamma_0^2-\Gamma_k(res)^2)^{0.5}$,
 where $\Gamma_k(res)$ is the momentum resolution, a Fermi step defined by $k_B T$ 
 and then convolve this with the instrumental resolution. 
 Both methods where tried with essentially identical results.
\end{thebibliography}
\end{document}